\input epsf

\documentstyle[prl,floats,aps]{revtex}

\tighten

\def\erf{\mathrm{erf}}

\begin{document}

\draft


\title{Evolution of Distorted Black Holes:  A Perturbative Approach}

\author{Gabrielle Allen${}^{(1)}$, Karen Camarda${}^{(4)}$, Edward
Seidel${}^{(1,2,3)}$}

\address{
$^{(1)}$Max-Planck-Institut f{\"u}r Gravitationsphysik,
Schlaatzweg 1, 14473 Potsdam, Germany \\
$^{(2)}$ National Center for Supercomputing Applications,
Beckman Institute, 405 N. Mathews Ave., Urbana, IL 61801 \\
$^{(3)}$ Departments of Astronomy and Physics,
University of Illinois, Urbana, IL 61801 \\
$^{(4)}$ Department of Astronomy and Astrophysics and Center for
Gravitational Physics and Geometry, \\
Pennsylvania State University, University Park, PA 16802
}

\date{\today}

\maketitle

\begin{abstract}
  We consider a series of distorted black hole initial data sets, and 
  develop techniques to evolve them using the linearized equations of 
  motion for the gravitational wave perturbations on a Schwarzschild 
  background.  We apply this to 2D and 3D distorted black hole 
  spacetimes.  In 2D, waveforms for different modes of the radiation 
  are presented, comparing full nonlinear evolutions for different 
  axisymmetric $\ell-$ modes with perturbative evolutions.  We show 
  how axisymmetric black hole codes solving the full, nonlinear 
  Einstein equations are capable of very accurate evolutions, and also 
  how these techniques aid in studying nonlinear effects.  In 3D we 
  show how the initial data for the perturbation equations can be 
  computed, and we compare with analytic solutions given from a 
  perturbative expansion of the initial value problem.  In addition to 
  exploring the physics of these distorted black hole data sets, in 
  particular allowing an exploration of linear, nonlinear, and mode 
  mixing effects, this approach provides an important testbed for any 
  fully nonlinear numerical code designed to evolve black hole 
  spacetimes in 2D or 3D.

\end{abstract}




\section{Introduction}
\label{sec:Introduction}

As numerical relativity is empowered by ever larger computers, 
numerical evolutions of black hole data sets are becoming more and 
more common\cite{Anninos96c,Seidel98b}.  The need for such simulations 
is great, especially as gravitational wave observatories like the 
LIGO/VIRGO/TAMA/GEO network are gearing up to collect gravitational 
wave data over the next decade (see, e.g., Ref.~\cite{Flanagan97b} and 
references therein).  A recent, thorough, and very detailed study 
indicates that black hole collisions are considered a most likely 
source of signals to be detected by these 
observatories\cite{Flanagan97b,Flanagan97a}.  Among the conclusions of 
this work is that reliable information about the merger process can be 
crucial not only to the interpretation of such observations, but also 
could greatly enhance the detection rate.  Therefore, it is crucial to 
have a detailed theoretical understanding of the coalescence process, 
and particularly the merger phase, that can only be achieved through 
numerical simulation.

However, numerical simulations of black holes have proved very 
difficult, largely because of the problems associated with dealing 
with singularities present inside.  Even in axisymmetry, at present it 
is difficult to evolve black hole systems beyond about $t=150M$, where 
$M$ is the mass of the system\cite{Anninos94b}. 
In 3D calculations, the huge memory requirements make these problems 
much 
more severe.  The most advanced 3D calculations based on traditional 
Cauchy evolution methods published to date, utilizing massively 
parallel computers, have difficulty evolving 
Schwarzschild~\cite{Anninos94c}, Misner~\cite{Anninos96c}, or 
distorted Schwarzschild~\cite{Camarda97b} beyond about $t=50M$.  
Characteristic evolution methods have been used to evolve distorted 
black holes in 3D indefinitely\cite{Gomez98a}, although to date 
waveforms have not been extracted and verified, and it is not clear 
whether the technique will be able to handle highly distorted or 
colliding black holes due to potential trouble with caustics.

In spite of these problems, much physics has been learned, especially 
in axisymmetry.  There, calculations of distorted black holes with 
\cite{Brandt94a,Brandt94b,Brandt94c} and without angular 
momentum\cite{Abrahams92a,Bernstein93b}, have been performed.  
Furthermore, calculations of the Misner two black hole initial data 
have been carried out\cite{Anninos94b,Anninos93b}, and the waveforms 
generated during the collision process have been extensively compared 
to calculations performed using perturbation 
theory\cite{Baker96a,Price94b,Pullin98a}.  One of the important 
results to emerge from these studies is that the numerical and 
perturbative results agree very well, giving great confidence in both 
approaches.  In particular, the perturbative approach turned out to 
work extremely well in some regimes where it was not, {\em a priori}, 
expected to be accurate.  This has led to considerable interest in 
comparing perturbative calculations to full scale numerical 
simulations, as a way of not only confirming the validity of each 
approach, but also as an aid to interpreting the physics of the 
systems\cite{Baker96a,Price94b,Pullin98a}.  We expect this rich interplay 
between perturbation theory and numerical simulation to play an 
important role in the future of numerical relativity.

To this end, in the present paper we study a family of distorted
axisymmetric and full 3D black hole initial data sets using
perturbative techniques.  These data sets consist of single black
holes that have been distorted by the addition of gravitational waves.
They can be considered to represent the final stages of the
coalescence of two black holes, just after the horizons have
merged.  They therefore provide an excellent system to study the late
stages of this process without having to model the long and difficult
inspiral period.

By using perturbative techniques to compute the waveforms expected in 
the evolution of these data sets, we provide an important testbed for 
full nonlinear codes that should evolve the same systems.  In regimes 
where the distortions are considered moderately small perturbations of 
the underlying Schwarzschild or Kerr geometries, full scale numerical 
calculations should agree well with the type of perturbative treatment 
presented here.  We expect that these results should have many uses in 
testing numerical relativity codes, and should also be useful in 
interpreting the physics contained in such simulations.  For cases 
where different 3D codes and techniques are being developed, these 
results should help certify and interpret the numerical results.  For 
example, 3D codes are being developed with standard ADM and new 
hyperbolic formulations of the equations\cite{Masso98a}, with 
different slicing conditions.  New approaches to black hole evolution 
that promise to extend dramatically the accuracy and duration of such 
calculations are under development, such as apparent horizon boundary 
conditions (see, e.g.,~\cite{Seidel92a,Cook97a} and references 
therein) or very recently characteristic 
evolution~\cite{Gomez98a,Gomez97a,Gomez97b}.  In all cases, these 
testbeds should be applicable and should provide important tests of 
the numerical simulation results.

\section{The Perturbative Approach to Evolution}
\label{sec:method}

In this section we provide the basic mathematical formalism for
evolving distorted black holes as perturbative systems.  We want to
investigate the possibility of treating the non-spherical multipole
moments of a general (numerical) spacetime as a linear perturbation
about its ``background'' spherical part.  Roughly speaking, this
approach should be a valid way to describe black hole spacetimes whose
nonspherical departure from Schwarzschild is small.  In practice, such
an approach has already shown itself to be spectacularly successful in
the ``close-limit'' approximation that regards the Misner initial data
for two colliding black holes as a nonspherical perturbation to
Schwarzschild~\cite{Pullin98a,Price94a}.  In this paper we apply similar
ideas to the evolution of distorted single black hole spacetimes,
providing detailed comparisons with fully nonlinear numerical
evolutions in axisymmetry, and providing the framework for comparing
fully 3D simulations with perturbative evolutions.  This approach is
also motivated by earlier work of Ref.~\cite{Abrahams92a}, where
perturbative evolution was used as a check on numerical evolution, but
not as a way to evolve Cauchy initial data.

We assume the spacetime metric $g_{\alpha\beta}$ to be in some general 
coordinate system $(t,\eta,\theta,\phi)$.  This metric comes either 
from an analytic solution to the Einstein equations, or a numerical 
simulation.  At this point we do not distinguish between these cases.  
In this coordinate system we can always write the metric in the form
$$
g_{\alpha\beta} = g^{sph}_{\alpha\beta} + h_{\alpha\beta}
$$
where $g^{sph}_{\alpha\beta}$ is the spherically symmetric $\ell=0$
term, in a decomposition of $g_{\alpha\beta}$ into tensor spherical
harmonics, and $h_{\alpha\beta}$ contains the higher multipoles which
describe the deviations from the spherically symmetric $\ell=0$ mode.
The terms $h_{\alpha\beta}$ satisfy, to ${\cal O}(h^2)$, 
the linear field equations,
linearized about $g^{sph}_{\alpha\beta}$.
Thus if the terms $h_{\alpha\beta}$ are small, in the sense that
$h_{\alpha\beta} \ll g^{sph}_{\alpha\beta}$, they can be well approximated
by a solution to the linearized equations.

Our general approach then, is to take Cauchy initial data which we
expect to describe a system which is close to spherical symmetry.
From this initial data, we use a decomposition into tensor spherical
harmonics to construct a background metric $g^{sph}_{\alpha \beta}$, and
perturbations $h_{\alpha\beta}$. The perturbations
$h_{\alpha\beta}$ are then evolved using the linear field equations.

The methods for extracting $g^{sph}_{\alpha\beta}$ and $h_{\alpha\beta}$,
and for evolving $h_{\alpha\beta}$ are now discussed in detail.

\subsection{Extraction}
\label{sec:extract}
We write the background metric in the form
\begin{equation}
  g^{sph}_{\alpha\beta} = \left(
\begin{array}{cccc}
-N^2(t,\eta)&0&0&0 \\
0&A^2(t,\eta) & 0 & 0 \\
0&0 & R^2(t,\eta) & 0 \\
0&0 & 0 & R^2(t,\eta) \sin^2\theta
\end{array}
\right)
\label{eqn:4dmetric}
\end{equation}
The perturbations $h_{\alpha\beta}$, which describe the
deviations from spherical symmetry, are expanded using Regge-Wheeler
harmonics~\cite{Regge57} as
\begin{eqnarray}
\label{eqn:rw1}
h_{tt} &=& -N^{2} H_0^{(\ell m)} Y_{\ell m} \\
h_{t\eta} &=& H_1^{(\ell m)} Y_{\ell m} \\
h_{t\theta} &=& h_0^{(\ell m)} Y_{\ell m,\theta}
 -c_0^{(\ell m)} Y_{\ell m ,\phi}/\sin\theta\\
h_{t\phi} &=& h_0^{(\ell m)} Y_{\ell m,\phi}
 -c_0^{(\ell m)} \sin\theta Y_{\ell m ,\theta}\\
h_{\eta\eta} &=& A^{2} H_2^{(\ell m)} Y_{\ell m} \\
h_{r\theta} &=& h_1^{(\ell m)} Y_{\ell m,\theta}
 - c_1^{(\ell m)}Y_{\ell m, \phi}/\sin\theta \\
h_{\eta\phi} &=& h_1^{(\ell m)} Y_{\ell m,\phi}
 + c_1^{(\ell m)} \sin\theta Y_{\ell m, \theta} \\
h_{\theta\theta} &=& R^2 K^{(\ell m)} Y_{\ell m} + R^2 G^{(\ell m)} Y_{\ell
m,\theta\theta} \nonumber\\
&&
+c_2^{(\ell m)} (Y_{\ell m, \theta\phi}-\cot\theta Y_{\ell
m,\phi})/\sin\theta \\
h_{\theta\phi} &=& R^2 G^{(\ell m)} \left( Y_{\ell m,\theta\phi} -
\cot\theta Y_{\ell m,\phi} \right) \nonumber\\
&&
-c_2^{(\ell m)} \sin\theta(Y_{\ell m,\theta\theta}-\cot\theta Y_{\ell m,\theta}
-Y_{\ell m}/\sin^2\theta)/2 \\
h_{\phi\phi} &=& R^2 K^{(\ell m)} \sin^2\theta Y_{\ell m} \nonumber\\
&& + R^2 G^{(\ell
m)} \left( Y_{\ell
    m,\phi\phi} + \sin\theta \cos\theta Y_{\ell m,\theta} \right)\nonumber\\
&&
\label{eqn:rw10}
-c_2^{(\ell m)}\sin\theta(Y_{\ell m,\theta\phi}-\cot\theta Y_{\ell m,\phi}).
\end{eqnarray}
In the above, there are seven even-parity $(H_0^{(\ell m)}$, 
$H_1^{(\ell m)}$, $h_0^{(\ell m)}$, $H_2^{(\ell m)}$, $h_1^{(\ell 
m)}$, $K^{(\ell m)}$, $G^{(\ell m)})$ and three odd-parity 
$(c_0^{(\ell m)},c_1^{(\ell m)},c_2^{(\ell m)})$ variables, which are 
functions only of $t$ and $\eta$.  In Eqs.~(\ref{eqn:rw1}) to 
(\ref{eqn:rw10}) a summation over the modes $\ell \ge 1$, $-\ell\le m\le \ell$ 
is understood.

Given this formal expansion of the full metric $g_{\alpha \beta}$, 
using the orthogonality of spherical harmonics we can extract the 
components of the background metric $g^{sph}_{\alpha \beta}$ by 
appropriate integrations over the 2-sphere
\begin{eqnarray}
  N^2 &=& - \frac{1}{4\pi}\int g_{tt} d\Omega
\label{eqn:lapse}
\\
  A^2 &=&  \frac{1}{4\pi}\int g_{\eta\eta} d\Omega
\\
  R^2 & = & \frac{1}{8\pi} \int \left(g_{\theta\theta}
   +\frac{g_{\phi\phi}}{\sin^2\theta}\right) d\Omega.
\end{eqnarray}
(We note that a similar expression in Ref.\cite{Abrahams92a} contained 
an error).  The ten variables describing the non-spherical 
contributions 
(for $\ell\ge 2$) can be extracted by similar integrals, 
for example
\begin{equation}
\label{eq:H2}
H_2^{(\ell m)}(t,\eta) = \frac{1}{A^2}  \int g_{\eta\eta} Y^*_{\ell m}
d\Omega.
\end{equation}
A complete list of formulae is given in 
Appendix~\ref{app:extraction_formulae}.  In the usual case where the 
full metric $g_{\alpha \beta}$ is given numerically, say after solving 
the Cauchy initial value problem of numerical relativity, these 
integrals over 2-spheres can be computed numerically.  In certain 
cases, the initial data may be known analytically, in which case the 
integrals can also be computed analytically.  In the sections below we 
shall encounter examples of both cases.  In any case, by performing 
these integrals over a series of 2-spheres of different radii on a 
given time slice (say, the initial time slice), we obtain the 
spherical background metric coefficients $N$, $A$, $R$ and for each 
nonspherical $\ell m-$mode the ten metric perturbation functions, 
e.g., $H_2^{(\ell m)}$.

These metric perturbation functions, coming directly from the metric, 
are gauge dependent; their values will depend on the particular gauge 
used, say, by the numerical code used to generate them.  However, a 
gauge-invariant perturbation theory of spherical spacetimes has been 
developed by various authors~\cite{Moncrief74,Gerlach79,Seidel90c}.  
As shown originally by 
Abrahams~\cite{Abrahams88,Abrahams89,Abrahams90}, such formalisms can 
be used in numerical relativity calculations to isolate the even- and 
odd-parity gauge-invariant gravitational wave functions.  The basic 
idea is that while the metric perturbation quantities, such as 
$H_2^{(\ell m)}$, will transform under infinitesimal coordinate 
transformations $x^{\mu} \rightarrow x^{\mu} + \delta x^{\mu}$ in the 
usual way, one can use this information to construct special 
quantities that are invariant under such gauge transformations.  On a 
Schwarzschild background these gauge-invariant quantities are found to 
obey the standard Regge-Wheeler and Zerilli equations describing 
gravitational waves propagating on the spherical black hole 
background.  We will follow this approach below.

To compute these gauge-invariant perturbations on the general, {\em 
time-dependent} background $g^{sph}_{\alpha\beta}$, we could follow 
the work of Refs.~\cite{Gerlach79,Seidel90c} and construct 
gauge-invariant functions from the extracted $\ell \ge 2$ multipole 
moments, and evolve these functions using the gauge-invariant 
linearized equations.  In this way we could treat data which is given 
in a general, and even time dependent, spherical coordinate system 
(although we note that in the present formalism the shift terms 
$g_{ti}$ must be treated as perturbations, i.e., they must be formally 
${\cal O}(h)$).  Here, we take a simpler approach, which has so far 
been suitable for our needs.  We assume that the Cauchy initial data 
for $g_{\alpha\beta}$ is, to ${\cal O}(h)$, given on a hypersurface of 
constant Schwarzschild time.  We then construct gauge invariant 
variables using Moncrief's prescription \cite{Moncrief74}, which can 
then be easily evolved using the Zerilli or Regge-Wheeler wave 
equations.  

We first transform $g_{\alpha\beta}$ to the areal radial 
coordinate, using
$$
r = R(t=0,\eta) = \sqrt{\frac{1}{8\pi} \int \left(g_{\theta\theta}
   +\frac{g_{\phi\phi}}{\sin^2\theta}\right) d\Omega},
$$
and then
$$
g_{rr} \approx \left(\frac{\partial R}{\partial \eta}\right)^{-2} g_{\eta\eta}.
$$
Note that $\partial R/\partial \eta$ can be easily calculated using
$$
\frac{\partial R}{\partial\eta} = \frac{1}{16\pi R}\int
    \left(\frac{\partial g_{\theta\theta}}{\partial \eta}
   +\frac{\partial g_{\phi\phi}}{\partial \eta}
    \frac{1}{\sin^2\theta}\right) d\Omega.
$$
Then, we calculate all the required multipole moments of this 
transformed metric, integrating the metric over 2-spheres as described 
above, using formulae detailed in the 
Appendix~\ref{app:extraction_formulae}.  This provides the ten metric 
perturbation functions as described above.

With this information, following Moncrief
we construct an odd-parity function $Q^\times_{\ell m}(t,r)$, and an
even-parity function $Q^+_{\ell m}(t,r)$, which are invariant under
first order coordinate transformations.  These are defined by
\begin{eqnarray}
 Q^{\times}_{lm}
  & = & \sqrt{\frac{2(l+2)!}{(l-2)!}}\left[c_1^{(\ell m)}
        + \frac{1}{2}\left(\partial_r c_2^{(\ell m)} - \frac{2}{r}
        c_2^{(\ell m)}\right)\right] \frac{S}{r}
\\
\text{and} \nonumber\\
Q^{+}_{lm}
&=&
\frac{1}{\Lambda}\sqrt{\frac{2(l-1)(l+2)}{l(l+1)}}
        \left(l(l+1)S
      (r^2\partial_r G^{(\ell m)}-2h_1^{(\ell m)})
\nonumber \right.
\\&&+\left.
        2r S(H_2^{(\ell m)}-r\partial_r K^{(\ell m)})+\Lambda r 
       K^{(\ell m)}\right).
\end{eqnarray}
Here we also note the definition
$$
S = 1-\frac{2M}{r},
$$
where $M$ is the Schwarzschild mass of the background. 
We approximate this mass by examining the function $m(r)$, given by
\begin{equation}
\label{eq:mass}
m(r) = \frac{r}{2}\left(1-\frac{1}{A^2}\right) \equiv M + {\cal O}(h).
\end{equation}
In practice, in the spacetimes we are considering here, the
function $m(r)$ is quite constant in $r$, and so there is no practical
ambiguity in computing the mass.  Alternatively, for very small
perturbations, one could consider simply using the ADM mass of the
spacetime for $M$, as is often done~\cite{Abrahams92a}, 
but this will include contributions from the waves in the spacetime.
Hence the mass defined by Eq.~(\ref{eq:mass}) provides a slight
improvement in the treatment.

While the individual metric quantities $h_{\alpha\beta}$ will
transform in the usual way by a linear gauge transformation $x^{\mu}
\rightarrow x^{\mu} + \delta x^{\mu}$, the functions $Q^{\times}_{lm}$
and $Q^{+}_{lm}$ are invariant, and thus are more directly connected
to the physics of the system.

We note that we have introduced a slight inconsistency in our
construction of the gauge-invariant quantities $Q^{\times}_{lm}$ and
$Q^{+}_{lm}$.  We have defined the Regge-Wheeler perturbation
functions (e.g. $H_2^{(\ell m)}$) in a general way, on a general
time-dependent background metric, but in computing the gauge-invariant
quantities we have simply assumed the background to be in
Schwarzschild coordinates.  More complex expressions for the
gauge-invariants on the more general background could be used, but in
the present applications this has not proved to be necessary.

At this stage we hint at a complication in this procedure which will we 
return to in more detail below.  The numerical prescription outlined 
above, using integrals over 2--spheres to pick off the different $\ell 
-m$ wave modes in a distorted black hole spacetime, simply lumps 
``everything not spherical'' into the wavefunctions.  It does not 
discriminate between contributions at various orders in the 
perturbation expansion, but rather it lumps them altogether.  
One must be careful when using the standard perturbation equations, 
as described below, which are derived from a formal theory keeping only
terms at linear order.  In particular, we will encounter cases where 
some perturbation modes appear {\em only} at higher order.

\subsection{Evolution}
\label{sec:evolution}
Having taken a general distorted black hole metric, we have detailed a
technique to compute the gauge-invariant functions describing the
waves on a Schwarzschild background.  We now turn to the linearized
evolution of these quantities.  These gauge invariant functions obey
the wave equations
\begin{eqnarray}
\label{eqn:rwwave}
  &&(\partial^2_t-\partial^2_{r^*})Q^\times_{lm}+ V_{RW}(r) Q^{\times}_{lm}
=  0
  \\
\label{eqn:zwave}
  &&(\partial^2_t-\partial^2_{r^*})Q^+_{lm}+ V_{Z}(r) Q^+_{lm}=0
\end{eqnarray}
where the Regge-Wheeler potential $V_{RW}(r)$ is given by
\begin{equation}
  V_{RW}(r) =
S\left[\frac{l(l+1)}{r^2}-\frac{6M}{r^3}
  \right]
\end{equation}
and the Zerilli potential is
\begin{eqnarray}
  V_{Z}(r) &=& S
   \left[
\frac{1}{\Lambda^2}\left(\frac{72M^3}{r^5}
-\frac{12M}{r^3}(l-1)(l+2)
  \left(1-\frac{3M}{r}\right)\right)\right.
\nonumber\\
&&\left.
 +\frac{l(l-1)(l+1)(l+2)}{r^2\Lambda}\right]
\end{eqnarray}
and finally $ \Lambda = (l-1)(l+2)+6M/r$ and $r^*= r+2M\ln(r/2M-1)$.

In general, the first time derivatives
of $Q^+_{\ell m}$ and $Q^\times_{\ell m}$ must also be calculated to
provide Cauchy data for the evolution.  In this paper only time
symmetric initial data has been used, for which
\begin{equation}
  \frac{\partial Q^+_{\ell m}}{\partial t} =
  \frac{\partial Q^\times_{\ell m}}{\partial t} = 0,
\end{equation}
but it is no problem to provide time derivatives through analysis of
the extrinsic curvature variables given as initial data in the more
general non-time symmetric case.

To summarize the development of this section, we have detailed the
technique we use to isolate the even- and odd-parity gauge-invariant
perturbation functions $Q^+_{\ell m}$ and $Q^\times_{\ell m}$ from a
numerically generated metric, under the assumption that the waves are
linear perturbations on a spherical background metric.  This
information can be used in two ways: (a) it can be used to extract
waveforms from a numerical simulation at a finite radius, and (b) it
can be used to provide initial data for the linearized evolution
equations given originally by Regge-Wheeler (odd-parity) and Zerilli
(even-parity).  These evolution equations can be used to compare
results obtained with the full nonlinear evolution of black hole
spacetimes.

\section{Application to Distorted Black Holes}

In this section we take the somewhat abstract discussion of the 
previous section and show how one applies it in practice to an actual 
family of numerically generated distorted black hole data sets.  The 
method is tested on the so-called black hole plus Brill wave 
spacetimes~\cite{Bernstein94a,Brandt97a}, which represent a black hole 
distorted by a gravitational wave.  The deviation from a Schwarzschild 
spacetime can be parameterized by a dimensionless parameter $a$, 
corresponding to the amplitude of the wave.  In the sections below we 
describe these initial data sets, describe tests we can perform on 
the procedures used to obtain initial data for the perturbation 
equations discussed above, and in Section~\ref{sec:evol} we
compare linear and nonlinear numerical evolutions.

\subsection{The Black Hole Initial Data}
\label{sec:initialdata}

In this section we review the single distorted black hole initial data
sets that we evolve in this paper.  These black holes are distorted by
the presence of an adjustable torus of nonlinear gravitational waves.
The amplitude and shape of the initial wave can be specified by hand,
as described below, and a range of initial data can be created, from
slightly perturbed to very highly distorted black holes. Such initial
data sets, and their evolutions in axisymmetry, have been studied
extensively, as described in
Refs.\cite{Abrahams92a,Bernstein93b,Bernstein94a}.  Three dimensional
versions have been developed and studied recently in
Refs.\cite{Camarda97b,Brandt97a,Camarda97a}.

Following\cite{Bernstein94a}, we write the 3--metric
in the form originally used by Brill~\cite{Brill59}:
\begin{equation}
\label{eq:metric}
d\ell^2 = \tilde{\psi}^4 \left( e^{2q} \left( d\eta^2 + d\theta^2 \right) +
  \sin^2\theta d\phi^2 \right),
\end{equation}
where $\eta$ is a radial coordinate defined by $\bar{r} = \frac{M}{2}
e^{\eta}$, where $M$ and $\bar{r}$ are the mass and standard
Schwarzschild isotropic radial coordinate of the black hole in the
Schwarzschild limit, described below.  In this paper, we choose our
initial slice to be time symmetric, so that the extrinsic curvature
vanishes, although more general datasets can also be
considered\cite{Brandt97a}.

Given a choice for the ``Brill wave'' function $q$, the Hamiltonian 
constraint leads to an elliptic equation for the conformal factor 
$\tilde{\psi}$.  The function $q$ represents the gravitational wave 
surrounding the black hole, and is chosen to be
\begin{equation}
\label{eq:q2d}
q\left(\eta,\theta,\phi\right) = a \sin^n\theta \left(
  e^{-\left(\frac{\eta+b}{w}\right)^2} +
  e^{-\left(\frac{\eta-b}{w}\right)^2} \right)
  \left(1+c \cos^2\phi\right).
\end{equation}
Thus, an initial data set is characterized by the parameters 
$\left(a,b,w,n,c\right)$, where, roughly speaking, $a$ is the 
amplitude of the Brill wave, $b$ is its radial location, $w$ its 
width, and $n$ and $c$ control its angular structure.  The parameters 
$n$, which must be a positive even integer, and $c$ dictate the 
angular ($\theta -\phi$) structure of the Brill wave.  In particular, 
if $c=0$, the resulting spacetime is axisymmetric.

A study of full 3D initial data and their evolutions are discussed 
elsewhere \cite{Brandt97a,Camarda97a,Camarda97c}.  If the amplitude $a$
vanishes, the undistorted Schwarzschild solution results, leading to
\begin{equation}
\tilde{\psi}^{(0)} = \sqrt{2M} \cosh \left( \frac{\eta}{2} \right).
\end{equation}

Note that the initial data defined by Eq.~(\ref{eq:metric}) and 
(\ref{eq:q2d}) have octant symmetry (i.e., equatorial plane symmetry 
and discrete symmetry in the four quadrants around the $z-$axis).  
Together with the time-symmetry condition, this implies that 
$Q^{\times}_{\ell m}(t,r) = 0$ and $Q^{+}_{\ell m}(t,r)$ is real, and 
only non-zero for even $\ell$ and $m$.  Therefore in the tests 
performed in this paper, only even-parity modes with even $\ell,m$ 
numbers are present, and in all analysis that follows linear 
evolutions will be carried out with the Zerilli evolution equation.  However, 
the techniques presented are quite general, and can be used on a wider 
class of initial data sets than considered here, containing the full 
range of modes.  All the data sets considered now, and in the 
following sections have $M=2$, $w=1$ and $b=0$, that is, loosely 
speaking, the Brill waves initially have unit width, and are centered 
on the throat of an $M=2$ Schwarzschild black hole.

The elliptic equation to be solved for $\tilde{\psi}$ must in general 
be solved numerically, as will normally be the case.  Once this is 
done, it can be evolved with a fully nonlinear numerical relativity 
code, and it can be also be used to compute initial data for the 
perturbation equations as described above.

However, in this case we can also write down a solution in terms of 
special functions which analytically solves the Hamiltonian constraint 
to linear order in the expansion parameter $a$.  This will provide a 
useful check on our procedures to compute linear initial data for the 
perturbation equations from the full nonlinear, numerically generated 
data.  We give these solutions in schematic form below, and in more 
detail in Appendix~\ref{app:linsol}.  

The conformal factor is expanded into spherical harmonics, such that
\begin{equation}
\tilde{\psi}(\eta,\theta,\phi) = \tilde{\psi}^{(0)}(\eta) 
+ a \sum_{\ell=0}^\infty \sum_{m=-\ell}^\ell 
  \tilde{\psi}^{(1)}_{\ell m}(\eta) Y_{\ell m}(\theta,\phi) 
+ {\cal O}(a^2)
\label{eqn:expansion}
\end{equation}
and the Hamiltonian constraint is solved to linear order in 
the expansion parameter $a$. Depending on the parameters
considered in Eq. (\ref{eq:q2d}), the resulting linear 
conformal factor has a different angular structure. Three
cases are considered in the following:

\begin{enumerate}
  
\item{$n=4$, $c=0$}: The only non-zero coefficients are 
    $\tilde{\psi}^{(1)}_{00}$, $\tilde{\psi}^{(1)}_{20}$ and 
    $\tilde{\psi}^{(1)}_{40}$.

\item{$n=2$, $c=0$}: In this axisymmetric case, the only non-zero 
 coefficients in the expansion are $\tilde{\psi}^{(1)}_{00}$ and 
  $\tilde{\psi}^{(1)}_{20}$.  The $\tilde{\psi}^{(1)}_{40}$ term that 
  one might expect is {\em missing}.  Note that this implies that at 
  linear order the $\ell=4$ Zerilli function vanishes, although it 
  will appear at order $a^{2}$.         
        
\item{$n=4$, $c\neq 0$}: In this non-axisymmetric case, the 
coefficients $\tilde{\psi}^{(1)}_{00}$, $\tilde{\psi}^{(1)}_{20}$, 
$\tilde{\psi}^{(1)}_{22}$, $\tilde{\psi}^{(1)}_{2,-2}$, 
$\tilde{\psi}^{(1)}_{40}$, $\tilde{\psi}^{(1)}_{42}$ and 
$\tilde{\psi}^{(1)}_{4,-2}$, are all non-zero.  In this case the 
$\tilde{\psi}^{(1)}_{44}$ contribution does not exist, and hence 
the $\ell=m=4$ Zerilli function will vanish at linear order.
\end{enumerate}

With these fully nonlinear numerical solutions and perturbative 
expansions we are in a position to extract and analyze initial data 
for the Zerilli evolution equation (and for the full 2D nonlinear 
evolution code), which is the topic of the next 
section.

\subsection{Extracting Wave Initial Data}

In this section we discuss specific examples of the extraction 
procedure applied to the axisymmetric black hole initial data sets 
discussed above.  Here we focus on the extraction of initial data 
itself, not on the evolution.  We use the exact analytic solution to 
the perturbative initial data equations as an aid to evaluating and 
understanding the numerically extracted initial data from the full 
nonlinear solution to the Hamiltonian constraint.

There are two main variations on the extraction of initial data that we 
compare and contrast in this section:

(a) We compute the gauge-invariant functions $Q^{+}_{\ell m}$ 
numerically, by computing numerical integrals over 2--spheres 
according to the procedure outlined above and in 
Appendix~\ref{app:extraction_formulae}.

(b) We can use the exact perturbative solution to calculate {\em
  analytically} the gauge-invariant variable denoted by
$Q^{+sph}_{\ell m}$, which we obtain by linearizing about the
spherically symmetric background, and then constructing the variables
as if the background was Schwarzschild.  This follows the spirit of
the general numerical procedure described above, and is the solution
to which we expect to converge with the complete numerical procedure.
This exact solution is too lengthy to be reproduced here, but can be
obtained straightforwardly with a computer algebra package.

We note that we can also use the exact solution to the perturbation
expansion to calculate a gauge-invariant variable, which results from
linearizing the spacetime about the background mass $M$ Schwarzschild
spacetime.  This is not quite the same as the wave $Q^{+sph}_{\ell m}$
obtained by linearizing about the more general spherical background.
This is due to two effects.  First, the individual metric
perturbation functions, such as defined via Eq.~(\ref{eq:H2}) above,
are computed on a more different spherical background than Schwarzschild
Second, the Brill wave introduces the presence of a
non-zero coefficients $\tilde{\psi}^{(1)}_{00}$ in the
expansion~(\ref{eqn:expansion}), which is reflected in the definition
of the mass defined in Eq.~(\ref{eq:mass}). The difference between the
two waveforms is ${\cal O}(a^2)$.

\subsubsection{Testing with Linear Solution}
\label{subsubsec:test}

The general method, and code, were tested using the exact linear
solutions for initial data using the conformal factor described above.
Using these exact linear solutions, the 3-metric components
(\ref{eq:metric}) were constructed on a 2D grid in $(\eta,\theta)$
coordinates (keeping only those terms linear in the Brill wave
parameter $a$).  The waveforms $Q^+_{20}$, $Q^+_{40}$ and $Q^+_{60}$
were numerically extracted, as described in Sec.~\ref{sec:extract},
and compared to the analytically computed functions $Q^{+sph}_{\ell
  m}$.  As expected from the numerical methods used, the numerically
extracted waveforms converged, as ${\cal O}(\Delta\theta^2)$, to
$Q^{+sph}_{\ell m}$, for each $\ell$-mode considered.  This provides
an excellent test of the accuracy of the integration routines on the
2--sphere required to compute the gauge-invariant wavefunctions, and
of the rather complex expressions involving the various
tensor-spherical harmonics for different $\ell$-modes that must be
coded.

\begin{figure}
\centerline{\mbox{\epsfysize=9cm
\epsffile{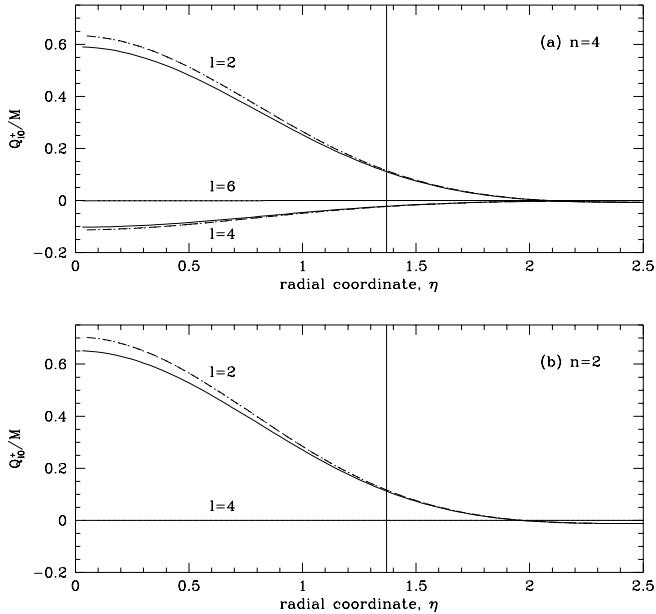}}}
\caption{Numerically extracted and exact initial data
  $Q^+_{\ell m}$ for the cases (a) $n=4$, $a=0.05$, $c=0$, and (b)
  $n=2$, $a=0.05$, $c=0$. In both graphs: the dashed lines show
  waveforms extracted exactly about the spherical background and the
  dotted lines waveforms extracted numerically from linear initial
  data. These are practically indistinguishable since they lie almost
  on top of each other. The solid lines are waveforms extracted
  numerically from nonlinear initial data.  The solid vertical line
  indicates the position of the peak of the background Schwarzschild
  potential in this radial coordinate. All the waveforms have been
  scaled by the appropriate extracted mass.  }
\label{extract1}
\end{figure}

As an example, Fig.~\ref{extract1} shows the numerically extracted and
exact waves for the initial data sets (a) $(a=0.05,n=4,c=0)$ and (b)
$(a=0.05,n=2,c=0)$, plotted as a function of the logarithmic radius
$\eta$.  The numerically and analytically computed wave, calculated
from the perturbative analytic solution to the Hamiltonian constraint,
are shown as the dotted and dashed lines in the figure. On this
scale they are almost indistinguishable, demonstrating the high
accuracy of the extraction procedure.  For comparison, we also show
the numerically extracted wave, discussed fully in the next section,
computed from the full nonlinear solution to the Hamiltonian
constraint, as a solid line. This indicates the difference between the
linear and nonlinearly computed initial data.

In Fig.~\ref{extract1}a, we show the $\ell=2,4,6$ modes.  For this
initial data set with $n=4$, the $\ell=2,4$ modes are present at
linear order, while the $\ell=6$ mode is not.  The numerically
computed mode from the linear solution should vanish, and does so at
second order in $\Delta \theta$. Correspondingly, for the $n=2$ linear 
initial data set, only the $\ell=2$ mode should be present, and this
is seen in Fig.~\ref{extract1}b, which shows the numerically extracted
$\ell =2$ mode to be indistinguishable from the exact result, and
the $\ell=4$ mode to be very close to zero.

\subsubsection{Extractions from the Nonlinear Initial Data}
\label{sec:extractnonlinear}

Now we use the procedure detailed in Sec.~\ref{sec:method} to extract
waveforms $Q^+_{\ell m}$ from (2D numerical) initial data generated by
solving the {\em nonlinear} constraints, for the same Brill wave
parameters as above in Sec.~\ref{subsubsec:test}.  In
Fig.~\ref{extract1} the initial waveform obtained is compared with the
previous waveforms from the exact linear initial data.  The waveforms
match very closely, with the greatest differences near the black hole
throat. This area, with the greatest difference between the waveforms
from the linear and nonlinear initial data, is also inside the maximum
of the background black hole potential, which is located (for $\ell=2$) 
at $\eta \approx 1.37$, and most of this part of the
waveform will be radiated across the horizon to disappear in the black
hole.

\begin{figure}
\centerline{\mbox{\epsfysize=9cm
\epsffile{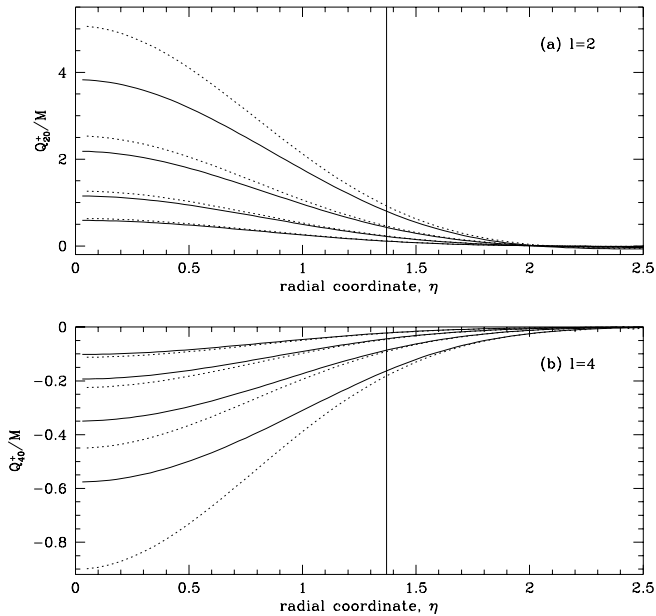}}}
\caption{
  Extracted waveforms from nonlinear initial data, compared to the
  corresponding exact linear solution $Q^{+sph}_{lm}$ for $n=4$, $c=0$
  for the different amplitudes $a=0.05$, $0.1$, $0.2$, and $0.4$. The
  solid lines show the waveforms extracted from the nonlinear initial
  data, and the dotted lines the exact linear waveforms.  All the
  waveforms have been scaled by the appropriate extracted mass, and
  increasing size of $a$ corresponds to increasing amplitude of the
  waveform. The solid vertical line indicates the position of the peak
  in the Schwarzschild potential.}
\label{extract2}
\end{figure}

\begin{figure}
\centerline{\mbox{\epsfysize=9cm
\epsffile{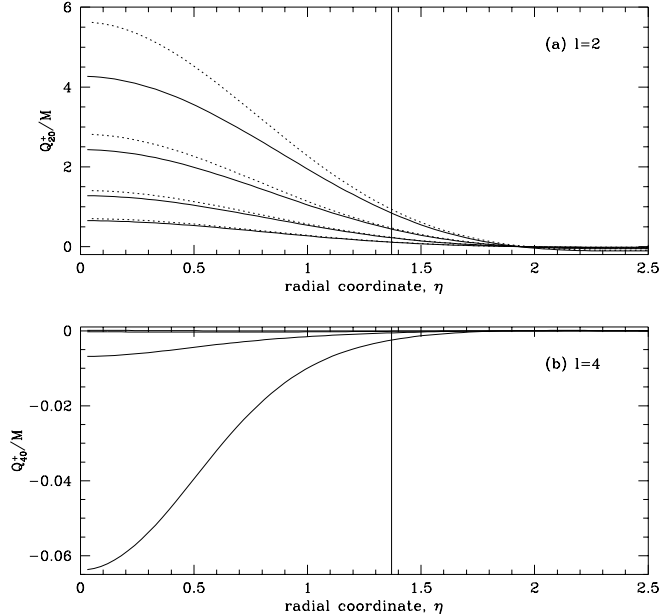}}}
\caption{
  Extracted waveforms from nonlinear initial data, compared to the
  corresponding exact linear solution $Q^{+sph}_{lm}$ for $n=2$, $c=0$
  for the different amplitudes $a=0.05$, $0.1$, $0.2$, and $0.4$. The
  solid lines show the waveforms extracted from the nonlinear initial
  data, and the dotted lines the exact linear waveforms.  All the
  waveforms have been scaled by the appropriate extracted mass, and
  increasing size of $a$ corresponds to increasing amplitude of the
  waveform.  Note that in the bottom figure, the amplitude of
  $Q^{+sph}_{40}$ is zero for all values of $a$. The solid vertical
  line indicates the position of the peak in the Schwarzschild
  potential.}
\label{extract3}
\end{figure}

As the Brill wave parameter $a$ is increased, the higher order terms
begin to have more influence, and the waveform obtained from the
nonlinear initial data deviates more from that from the linear initial
data.  This is demonstrated in Figs.~\ref{extract2} and
\ref{extract3}, in which $Q^+_{\ell m}$ as extracted from nonlinear 
data is shown for a range of Brill wave amplitudes, $a$. The Figures
show that in the region close to the black hole throat (at $\eta =0$) the
amplitude of $Q^+_{\ell m}$ does not increase linearly with $a$, as expected.
However, at the maximum of the black hole potential, $\eta\approx
1.37$ (for $\ell=2$ and $M=2$), $Q^+_{\ell m}$  
scales nearly linearly with $a$.

We will see in the next section that even in cases such as these where 
the linear study deviates significantly from the nonlinear extraction, 
the linear evolution can still do a reasonably good job in predicting the 
results of the full nonlinear evolution far from the black hole.  The 
key reason for this seems to be that the nonlinearities of these 
particular datasets are largest well inside the peak of the potential 
barrier, and hence they are kept inside and propagate down the black 
hole.  This point will be discussed further in the next section.

In this section we have demonstrated that our extraction method is 
accurate by comparisons with known linear solutions, and shown that 
the nonlinear initial data sets, parameterized by the 
Brill wave parameter $a$, yield initial waveforms which agree 
closely with the known linear solution for small $a$.
In the next section we turn to evolutions of these numerically extracted
linear initial data sets.

\section{Evolution of Axisymmetric Initial Data sets}
\label{sec:evol}

The waveforms $Q^+_{\ell m}$, extracted from data on the initial
hypersurface, can be numerically evolved using the Zerilli
equation~(\ref{eqn:rwwave}).  We stress that in this case, we compute
$Q^+_{\ell m}$ from the {\em nonlinear} computed initial data, {\it
i.e.} from the full solution to the Hamiltonian constraint. In the
absence of an exact solution for the perturbative initial data
expansion, as will normally be the case, this will be the only way to
obtain initial data for the linearized evolution equations.  For the
cases considered in this paper, only time symmetric initial data was
considered, with $\dot{Q}^+_{\ell m}=0$ on the initial hypersurface.
For non-time symmetric data, $\dot{Q}^+_{\ell m}$ can be calculated
using the numerical extrinsic curvature, using a procedure similar to
that for extracting $Q^+_{lm}$ from $g_{ij}$ \cite{Abrahams95b}.

In the following sections, we compare these 1D linear evolutions, with
the results from 2D fully nonlinear evolutions of the original initial
data sets \cite{Bernstein93b,Bernstein94a}. The data is compared by constructing,
in both simulations, $Q^+_{\ell m}(t)$ at the same Schwarzschild
radius. In the nonlinear code $Q^+_{\ell m}(t)$ this requires the same
extraction procedure that was used to find the linear initial data, to
be applied at the chosen Schwarzschild radius throughout the
evolution. In all the examples shown here, the radiation 
waveforms were extracted at a Schwarzschild radius $r=15M$. 

In principle one must be careful in comparing waveforms measured with 
different time coordinates.  Since the nonlinear evolution implements 
a general slicing condition (in these examples an algebraic slicing), 
and not the Schwarzschild slicing used in the linear evolution, it 
could happen that corrections would be needed to account for the 
differences in slicing.  However, in our simulations the slicings are 
similar enough in the regions where the waves are computed that this 
is only a very small effect, as is borne 
out by the Figures.

For all the results shown here, both for the linear and nonlinear
evolutions, the accuracy of the results were carefully studied, to be
sure that any differences between the nonlinear and linear waveforms
is due only to the different initial data or the mode of evolution.
This is particularly important to ensure that the observed modes are
not due to insufficient resolution or boundary effects.

\subsection{\bf Evolutions for n=4, c=0}
 
First, we discuss the evolutions corresponding to the axisymmetric
initial data sets specified by $n=4$, $c=0$, for a range of Brill wave
amplitudes $a$. For the exact linear initial data described in 
 Sec.~\ref{sec:initialdata}, it was seen that $Q^+_{20}$ and $Q^+_{40}$ 
occurred to linear order in the Brill wave amplitude $a$, with all other modes
occurring at higher order. This linear scaling of $Q^+_{20}$ and $Q^+_{40}$
was also seen in Fig.~\ref{extract2}, in
the initial data used for these linear evolutions. 

\begin{figure}
\centerline{\mbox{\epsfysize=9cm
\epsffile{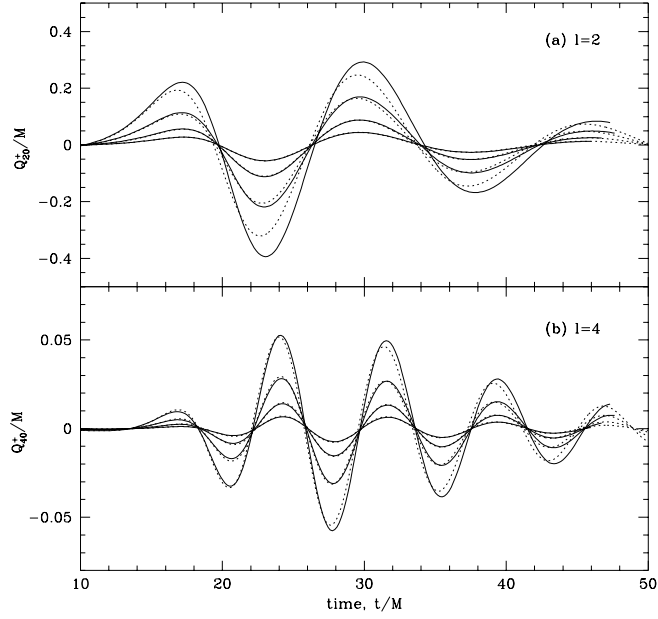}}}
\caption{
\label{n4_compare}
Comparison of $\ell=2$ and $\ell=4$ waveforms at $r=15M$ from linear
(dotted line) evolutions of linearized initial data and nonlinear
(solid line) evolutions of nonlinear initial data. The initial data
were constructed with the Brill wave parameters $n=4$, $c=0$, for four
amplitudes, $a=0.05$, $0.1$, $0.2$ and $0.4$.}
\end{figure}

Fig.~\ref{n4_compare} compares the radiation 
waveforms at $r=15M$  from the linear evolution of the Zerilli 
equation and from the nonlinear evolution of the full Einstein field 
equations.  It is clearly seen that for the lowest amplitude Brill 
waves, the waveforms from linear and nonlinear evolutions match very 
closely. As the amplitude increases, the deviations become more 
apparent.

\begin{figure}
\centerline{\mbox{\epsfysize=9cm
\epsffile{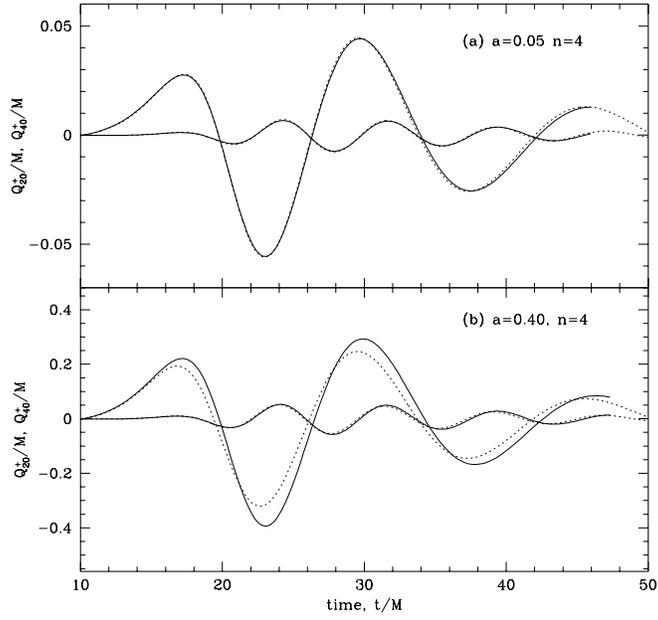}}}
\caption{
\label{lowhigh_n4}
Comparison of $\ell=2$ and $\ell=4$ waveforms at $r=15M$ from linear
(dotted line) evolutions of linearized initial data and nonlinear
(solid line) evolutions of nonlinear initial data. Figure (a)
corresponds to the initial data set $n=4$, $c=0$ and $a=0.05$, and
Figure (b) corresponds to the higher amplitude case $n=4$, $c=0$ and
$a=0.4$.}
\end{figure}

In Fig.~\ref{lowhigh_n4} we isolate the cases $a=0.05$ and $a=0.4$ for
closer study, and compare the relative amplitudes of the $\ell=2$ and
$\ell=4$ waveforms. This clearly shows the high level of agreement for
the lowest amplitude, shown in Fig.~\ref{lowhigh_n4}.  Note that the
scales have been chosen in the two graphs, such that if the waves were
scaled linearly with $a$, they would be the same size in both (a) and
(b). Although we have {\em evolved} the same (non-linear) initial data
in both cases, nonlinearities are clearly coming in, even in the 
initial data.

\subsection{\bf Evolutions for n=2, c=0} 

The second group of initial data sets were created using
$n=2$ and $c=0$, for the same range of Brill wave amplitudes
as in the previous section. For these parameters, the exact
linear initial data sets of Sec.~\ref{sec:initialdata} contained
only the $Q^+_{20}$ mode to linear order in $a$, with all other
modes occurring at second order or higher in $a$. This behavior
is also seen in Fig.~\ref{extract3}, in the initial data used 
for these linear evolutions.

\begin{figure}
\centerline{\mbox{\epsfysize=9cm
\epsffile{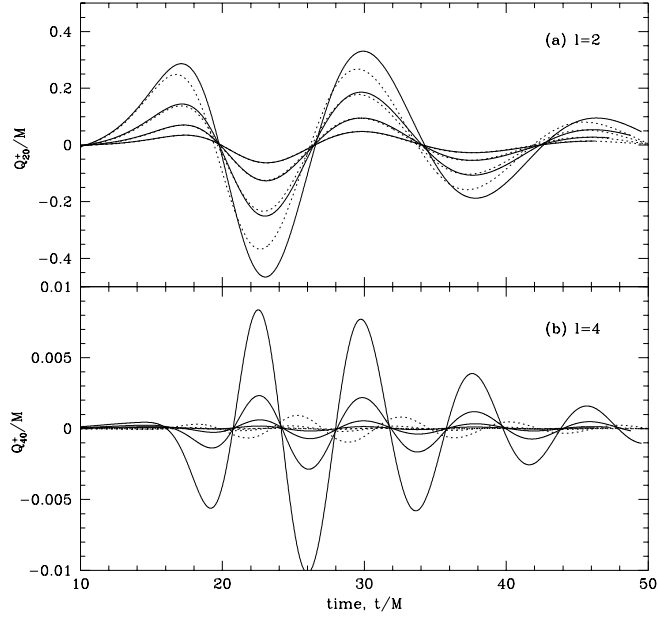}}}
\caption{
\label{n2_compare}
Comparison of $\ell=2$ and $\ell=4$ waveforms at $r=15M$ from (a) linear
(dotted line) evolutions of linearized initial data and (b) nonlinear
(solid line) evolutions of nonlinear initial data. The initial data
was constructed with the Brill wave parameters $n=2$, $c=0$, for four
amplitudes, $a=0.05$, $0.1$, $0.2$ and $0.4$.}
\end{figure}

Fig.~\ref{n2_compare} compares the radiation waveforms at $r=15M$ from 
the linear evolution of the Zerilli equation and from the nonlinear 
evolution of the full Einstein field equations.  For these initial 
data, the $\ell=2$ waveforms shown in Fig.~\ref{n2_compare}a again 
match well for low amplitudes, with deviations growing for the higher 
amplitudes.  The $\ell=4$ waveforms, shown in Fig.~\ref{n2_compare}b 
are of much lower amplitude than the $\ell=2$ waveforms.  As is 
apparent in the Figure, these waveforms from the linear and nonlinear 
evolutions do not match well, in amplitude or phase for any of the 
Brill wave parameters used.  However, we should not expect them to.  
We have already seen in Sec.~\ref{sec:initialdata}, in these $n=2$, 
$c=0$ initial data sets, there is {\em no} $\ell=4$ content at linear 
order in the expansion parameter $a$.  The expansion of the initial 
data in powers of the amplitude $a$ is directly related to the formal 
expansion used in the derivation of the Zerilli evolution equation.  
The equation is valid only for evolving initial data of order $a$, but 
not order $a^2$, since all such terms were dropped in the derivation.  
In the case in point, the $\ell =4$ data are of order $a^2$ and hence 
this data will not be accurately evolved with the linear evolution 
equation.  Basically this means that the $\ell =4 $ piece, being of 
order $a^2$, is {\em too small} to be treated by linear theory, and 
only the nonlinear code can pick this up.

As shown by Gleiser {\it et al}, \cite{gleiser96a}, the Zerilli 
equation can be generalized to higher order, in which case it has the 
same form as for the linear case, but now with nonlinear source terms 
that mix in different multipoles.  In this case we could in principle 
evolve the $\ell = 4$ data with such an equation, taking into account 
quadratic terms in the linear $\ell =2$ (and other) modes.  In 
practice however, this would not be possible with the extraction 
method as used here, since errors of order $a^2$ are already 
introduced.  In fact, it is worth stressing that the standard linear 
extraction process used to obtain the waveforms for the nonlinear 
code has been applied even for these nonlinear modes.  The 
implications of such higher order effects on waveform extraction 
should be carefully considered in future investigations.

The main point to be made here is that linear theory does not, and 
should not agree with the nonlinear code in this case, and one must be 
careful in applying linear theory as a testbed.  On the other hand, 
these techniques open the door to careful and systematic studies of 
nonlinear physics of black holes, such as mode mixing, that could be 
pursued in the future.

\begin{figure}
\centerline{\mbox{\epsfysize=9cm
\epsffile{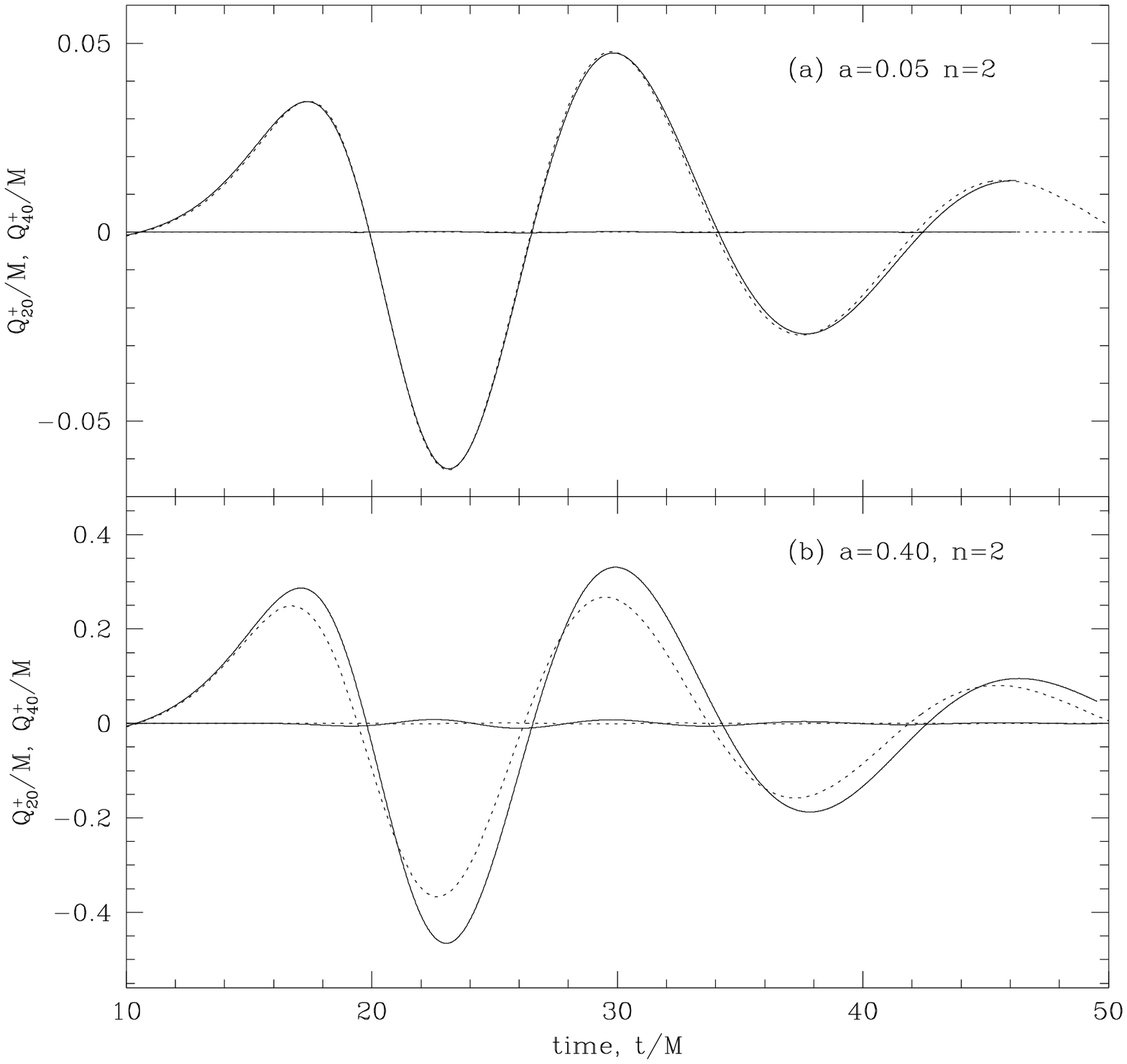}}}
\caption{
\label{lowhigh_n2}
Comparison of $\ell=2$ and $\ell=4$ waveforms at r=15M from linear
(dotted line) evolutions of linearized initial data and nonlinear
(solid line) evolutions of nonlinear initial data. Figure a) 
corresponds to the initial data set $n=2$, $c=0$ and $a=0.05$, and
Figure b) corresponds to the higher amplitude case $n=2$, $c=0$ and $a=0.4$.}
\end{figure}

Fig.~\ref{lowhigh_n2} shows the $\ell=2$ and $\ell=4$ waveforms on the
same graph, for the low amplitude $a=0.05$ and higher amplitude
$a=0.40$ cases. Fig.~\ref{lowhigh_n2}a shows results from the $a=0.05$
initial data, showing very good agreement between the linear and
nonlinear waveforms. The $\ell=4$ waveforms are too small to be seen
on this scale. Fig.~\ref{lowhigh_n2}b shows results from the higher
amplitude $a=0.40$ case. Here the discrepancy between the two methods
can be seen in both the $\ell =2$ and $\ell =4$ waveforms.

One might expect, given the rather large discrepancy between linear 
and nonlinear initial data for the $a=0.40$ case, that the linear and 
nonlinear evolved waveforms would be rather more different than they 
are.  However, as previously noted the largest deviations between the 
linear and nonlinear initial data occurs near the horizon, well inside 
the peak of the potential barrier.  Although linear theory breaks down 
inside the potential barrier, linear theory is still fairly accurate 
outside the peak where waveforms are measured.  This effect has also 
been seen in studies of black hole collisions (see, e.g. 
\cite{Anninos94b,Anninos93b,Baker96a,Price94b,Price94a,Gleiser96b}) 
where perturbative treatment was very successful even when one might 
naively think it would break down.

\subsection{Radiated Energy}

The asymptotic radiated energy flux, for each
angular mode, can be defined by \cite{Abrahams95b}
\begin{equation}
  \frac{dE^{\ell m}}{dt} = \frac{1}{32\pi} \left(
  \left|\frac{\partial Q^{+}_{\ell m}}{\partial t}\right|^2
 +\left|\frac{\partial Q^{\times}_{\ell m}}{\partial t}\right|^2
  \right)
\end{equation}
which can easily be numerically integrated to give the radiated
energy, $E^{lm}$ in each mode. Note that this energy formula comes
from {\em linear} theory, and hence will not provide an accurate 
prediction for the energy in the higher order modes. 

\begin{figure}
\centerline{\mbox{\epsfysize=9cm
\epsffile{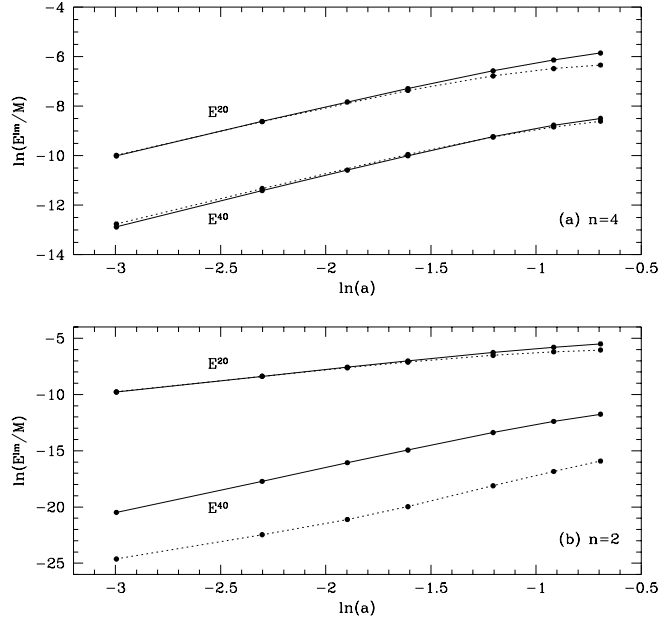}}}
\caption{
\label{energy}
The radiated energy, calculated at $r=15M$, using the radiation
waveforms from both the linear (dashed lines) and nonlinear (solid
lines) evolutions. The top graph is for the data set $n=4$, $c=0$, and
the bottom graph is for the data set $n=2$, $c=0$. In both cases, the
range of Brill wave amplitudes $a=0.05$ to $a=0.40$ was used.}
\end{figure}

Fig.~\ref{energy} uses log plots to compare the energies in the
waveforms from the nonlinear and linear evolutions of the two initial
data sets described above.  For the $n=4$ initial data sets, shown in 
Fig.~\ref{energy}a, the
energy radiated in both $\ell=2$ and $\ell=4$ waveforms is similar, for the
two types of evolution and for the range of Brill wave amplitudes
considered. The energy carried by the $\ell=4$ waveform is some 10 to 20
times smaller than the energy carried by the $\ell=2$ waveform. 
The radiated energy scales as ${\cal O}(a^2)$, for both modes.

In Fig.~\ref{energy}b we show the $n=2$ family of initial data sets,
and the picture is different, as was already seen in the
evolutions. Whereas the radiated energy is similar from both methods
of evolution for the $\ell=2$ waveform, the $\ell=4$ waveform is very
different. The $\ell=4$ waveform resulting from the nonlinear
evolution, contains a factor of around 60 times more energy than the
corresponding waveform from the linear evolution. The ``radiated energy''
in the $\ell=4$ waveform scales as ${\cal O}(a^4)$, for both methods
of evolution. But we emphasize that we have used a linearized energy measure
in this case, where nonlinear effects should be accounted for.

\section{3D Calculations}

So far in this paper we have concentrated on axisymmetric initial data 
sets, in order to aid the understanding and interpretation of the 
method.  In this section, the same procedure is applied to construct 
linearized initial data from a non-axisymmetric 3D initial data set.  
Our aim here is to show that the extraction and linear analysis 
techniques developed here carry over easily to the full 3D case, even 
though the extraction process is more complicated.  The evolutions of 
these data sets, and comparisons with nonlinear calculations, are left 
for the next paper in this series~\cite{Allen98a}.

The 3D initial data set studied here again corresponds to a black hole 
distorted by a gravitational wave, belonging to the same family as
described in Sec.~\ref{sec:initialdata}. The non-axisymmetry follows from
choosing a non-zero parameter $c$ in Eq.~\ref{eq:q2d}. The nonlinear
data was created using the parameters $a=-0.1$, $c=0.5$, $n=4$, $b=0$
and $w=1$. 

Again, an exact linear solution for the 3-metric was constructed for these
parameters, as detailed in Appendix~\ref{app:linsol}. Starting with 
this exact  3-metric on the initial hypersurface, exact expressions 
for the waveforms $Q^{+sph}_{\ell m}$ were found, using the 
extraction procedure of Sec.~\ref{sec:method}. 
Using the exact linear solutions, the 3-metric was calculated
numerically, on a 3D polar grid, in $(\eta,\theta,\phi)$ coordinates
(keeping only those terms linear in the Brill wave parameter $a$). 
Waveforms for different $\ell$ and $m$ were then numerically extracted
and compared to the exact $Q^{+sph}_{\ell m}$. As before, the numerically
calculated waveforms converge to the exact waveforms as ${\cal O}(\Delta
\theta^2, \Delta \phi^2)$ as expected from the numerical methods.

\begin{figure}
\centerline{\mbox{\epsfysize=9cm
\epsffile{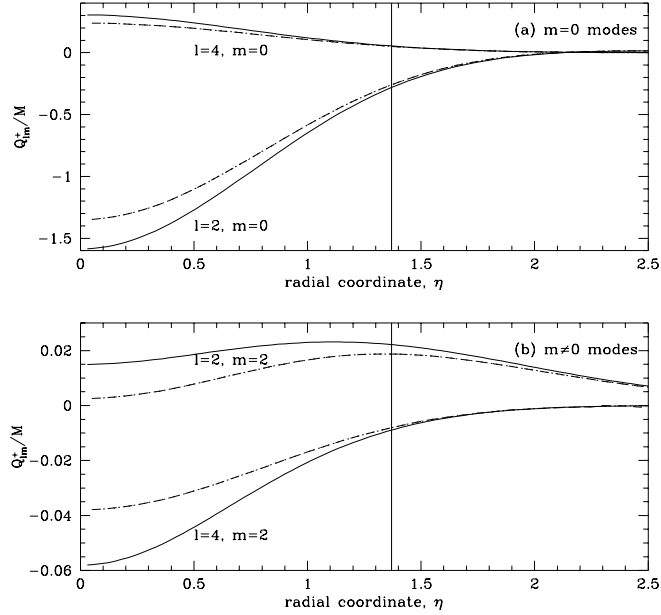}}}
\caption{Numerically extracted and exact initial data
  $Q^+_{\ell m}$ for the case $n=4$, $a=-0.1$, $c=0.5$. 
  The top and bottom show respectively the $m=0$ and $m\neq 0$
  waveforms which appear in the
  exact linear solution. 
  In both graphs: the short dashed lines show waveforms extracted 
  exactly about the spherical
  background; the dotted line waveforms extracted numerically from
  linear initial data; and solid lines waveforms extracted numerically
  from nonlinear initial data.  All the waveforms have been scaled by
  the appropriate extracted mass, and the solid vertical lines indicate
the location of the peak in the Schwarzschild potential barrier.  }
\label{ext3D_1}
\end{figure}

Fig.~\ref{ext3D_1} shows the numerically extracted and exact 
waveforms, for the modes which are present to ${\cal O}(a)$ in the 
exact solution.  In this figure, the top graph shows the higher 
amplitude axisymmetric $(m=0)$ modes, and the bottom graph the lower 
amplitude non-axisymmetric $(m \neq 0)$ modes.  The two solutions, 
shown by the dashed and dotted lines, cannot be distinguished for this 
resolution, ($N_\eta=200$,$N_\theta=52$,$N_\phi=52$).

Finally, instead of extracting the waveforms from the numerical linear 
initial data, we extract from 3D numerical data, found by solving the 
nonlinear constraints for the given Brill wave parameters.  This 
`nonlinear' initial data is constructed in the $(\eta,\theta,\phi)$ 
coordinate system.  The waveforms obtained from this initial data are 
shown as solid lines in Fig.~\ref{ext3D_1}.  Note that there are 
significant differences between the waveforms extracted from the 
linear and initial data.  However, the largest differences are again 
inside the maximum of the potential barrier of the background 
spacetime at $\eta \approx 1.37$ for $\ell=2$.  On the potential 
barrier the linear and nonlinear waveforms are very similar, with the 
largest difference occurring for $\ell=2, m=2$, indicating that in an 
evolution of this time-symmetric initial data, there should be only 
small differences outside of this radius.  Although not shown here, 
the difference between the initial waveforms from the linear and 
nonlinear data decreases as the Brill wave parameter $a$ is reduced.

\begin{figure}
\centerline{\mbox{\epsfysize=4.5cm
\epsffile{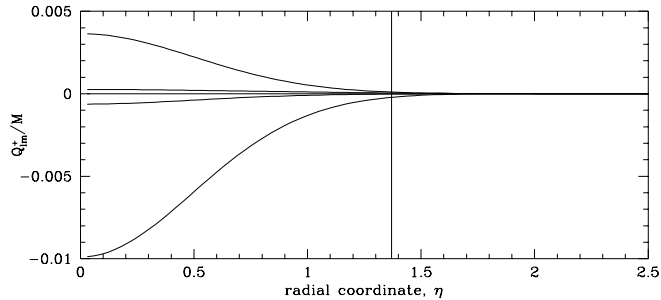}}}
\caption{Numerically extracted $Q^+_{\ell m}$ for the case 
  $n=4$, $a=-0.1$, $c=0.5$. The graph shows waveforms extracted
  from the nonlinear initial data, for those $\ell$ and $m$ modes
  which are not present in the exact linear solution. From 
  top to bottom, the waveforms are $Q^+_{62}/M$, $Q^+_{44}/M$, 
  $Q^+_{66}/M$, $Q^+_{64}/M$ and $Q^+_{60}/M$. Here $M$ is the
  extracted mass.}
\label{ext3D_1b}
\end{figure}

Fig.~\ref{ext3D_1b} shows the initial waveforms which are present 
in the extraction from nonlinear data, but are not found in the 
linear initial data, where they occur at ${\cal O}(a^2)$ or higher.
As in the axisymmetric cases discussed above, these modes should
{\em not} agree with full nonlinear evolutions.

\section{Conclusions}
\label{sec:conclusions}

Building on previous work 
(e.g.,~\cite{Abrahams92a,Price94a,Abrahams95b}) we have further 
developed the use of perturbation theory as a tool for numerical 
relativity.  We presented techniques for studying single, distorted 
black holes without net angular momentum, using the perturbative 
Regge-Wheeler and Zerilli equations.  These techniques include 
extracting initial data, decomposed into different $\ell,m$ modes, 
from a numerically computed black hole initial data set, and evolving 
these modes forward in time with the perturbation equations.  These 
results can then be used both to perform a powerful check on the 
accuracy of fully nonlinear evolution codes, from which waveforms can 
be extracted, and to help untangle linear and nonlinear effects, such 
as mode-mixing, of the full evolution.

We applied this technique to a series of distorted black holes, using 
the ``Brill wave plus black hole'' family of initial data 
sets~\cite{Bernstein94a,Brandt97a}.  These data sets have adjustable parameters 
for the shape and amplitude of the initial distortions, and have been 
well studied previously and found to mimic the behavior of two black 
                        holes that have just collided head-on.  They have a further nice 
property that an analytic solution to the initial value problem is 
possible if the Brill wave amplitude is small.  This information was 
used both to test the accuracy of the perturbative initial data 
extraction techniques, and to understand the spectral decomposition of 
both the linear and {\em nonlinear} contributions to the initial data.

We then performed a series of axisymmetric, fully nonlinear 
supercomputer evolutions of the full initial data sets, and compared 
gravitational radiation waveforms extracted from these simulations to 
results obtained via our perturbative techniques using the Zerilli 
equation.  We find that for initial data sets containing $\ell=2,4$ 
modes that can be shown to be {\em linear} in the wave amplitude 
parameter, the extracted waveforms from the full nonlinear evolutions 
show excellent agreement with the perturbatively evolved waveforms, 
confirming both the very high degree of accuracy that can be achieved 
with axisymmetric codes and the linear perturbation code and approach 
to the problem.

However, we also find that some modes computed with the nonlinear 
simulation code do {\em not} agree with the linearized treatment; in 
some cases the linear evolution can give very different results from 
the full nonlinear evolution for specific modes, even when other modes 
agree well.  But in all such cases, one can see from analytic study of 
the initial data that it is precisely these ``renegade'' modes that 
show up at a {\em nonlinear} order in the wave amplitude parameter.  
Their perturbative evolution equations should then be extended to 
second order, with nonlinear source terms (composed of linear modes) 
coming in.  In effect, such modes are {\em too small} to be treated by 
linear theory, since nonlinear contributions from other modes are 
relatively large enough not to be neglected.  Hence one can see 
clearly the effects of mode-mode coupling in these cases.  The 
combination of linearized treatment and perturbative treatment was 
essential not only to confirm the nonlinear code, but also to 
interpret the complex nature of the numerical results.  Furthermore, 
we pointed out that the use of linear waveform extraction techniques 
should be studied further for systems that may contain modes only at 
higher, nonlinear order.

Another important point to mention is the role that the potential 
barrier plays in the evolution of these distorted black holes.  The 
application of perturbation theory to the evolution of these distorted 
black holes often works well, even in cases where one can see large 
deviation between linear and nonlinear treatment in the initial 
data.  But in such cases, the error made by using a linear 
approximation is largest {\em inside} the peak of the potential 
barrier, and hence most of this nonlinearity is swallowed by the black 
hole.  Hence, as has also been stressed in previous work~\cite{Price94b}, 
the potential barrier acts to trap much of the deviation from linear 
behavior, extending the range of applicability of perturbative 
treatment over what one might naively expect.

Finally, we also applied this technique to study a new class of 3D 
distorted black hole initial data sets~\cite{Brandt97a}, and we 
showed that the initial data can be extracted accurately for study in 
3D, just as in axisymmetry.  The comparison of evolutions of such 3D 
distorted black hole initial data between full nonlinear numerical 
relativity and perturbation theory will be the subject of the next 
paper in this series.

In this paper we have only considered examples of time symmetric, 
even-parity perturbations of non-rotating black holes.  The technique 
is much more general, and also applies to all even- and odd-parity 
modes, and to non-time symmetric initial data, which will be 
considered in future papers in this series.  

The data sets developed in Ref.\cite{Brandt97a} also contain distorted 
black hole initial data with angular momentum.  The rotating case is 
considerably more complicated, and naturally involves using the 
Teukolsky equation to evolve perturbations of the Kerr metric.  This 
important followup step will be considered in a future paper.

\acknowledgements This work has been supported by AEI, NCSA, and the 
Binary Black Hole Grand Challenge Alliance, NSF PHY/ASC 9318152 (ARPA 
supplemented).  We thank S. Brandt for help with the analytic solutions
for the perturbative initial data, and H. Beyer, C. Gundlach and K. Kokkotas 
for helpful discussions. E.S. would like to thank J. Mass\'o and C. Bona for 
hospitality and discussions at the University of the Balearic Islands 
where part of this work was carried out.  Calculations were performed 
at AEI and NCSA on an SGI/Cray Origin 2000 supercomputer.

\appendix

\section{Extraction Formulae}
\label{app:extraction_formulae}
Here we list the general formulae for extracting the Regge-Wheeler variables
from the metric $g_{\alpha\beta}$ using the 
 the spherical background $g^{sph}_{\alpha\beta}=
\mbox{diag}(-N(t,r)^2,A(t,r)^2,R(t,r)^2,R(t,r)^2\sin^2\theta) $,
\begin{eqnarray}
c_0^{(\ell m)}(t,r) &=& \frac{1}{\ell(\ell+1)}
        \int \frac{1}{\sin\theta}(g_{t\phi}Y^*_{\ell m,\theta}
        -g_{t\theta}Y^*_{\ell m,\phi}) d\Omega
\\
c_1^{(\ell m)}(t,r) &=& \frac{1}{\ell(\ell+1)}
        \int \frac{1}{\sin\theta}(g_{r\phi}Y^*_{\ell m,\theta}
        -g_{r\theta}Y^*_{\ell m,\phi}) d\Omega
\\
c_2^{(lm)}(t,r) &=& -\frac{2}{\ell (\ell+1)(\ell-1)(\ell+2)}
        \int\left\{\left(-\frac{1}{\sin^2\theta}g_{\theta\theta}+\frac{1}
        {\sin^4\theta}g_{\phi\phi}\right) (\sin\theta Y^*_{\ell
        m,\theta\phi}-\c Y^*_{\ell m,\phi})\right.
\\
        &&\left.+ \frac{1}{\sin\theta} g_{\theta\phi} (Y^*_{\ell
           m,\theta\theta}-\cot\theta Y^*_{\ell m,\theta}
           -\frac{1}{\sin^2\theta}Y^*_{\ell m,\phi\phi})
           \right\}d\Omega
\\
h_0^{(\ell m)}(t,r) &=&  \frac{1}{\ell (l+1)} \int \left\{ g_{t\theta}
            Y^*_{\ell m,\theta} + \frac{1}{\sin^2\theta}
            g_{t\phi}Y^*_{\ell m,\phi}\right\} d\Omega
\\
h_1^{(\ell m)}(t,r) &=& \frac{1}{\ell (l+1)} \int \left\{ g_{r\theta}
            Y^*_{\ell m,\theta} + \frac{1}{\sin^2\theta}
            g_{r\phi}Y^*_{\ell m,\phi}\right\} d\Omega
\\
H_0^{(\ell m)}(t,r) &=& \frac{1}{N^2}\int g_{tt}Y^*_{\ell m}d\Omega
\\
H_1^{(\ell m)}(t,r) &=& \int g_{tr} Y^*_{\ell m} d\Omega
\\
H_2^{(\ell m)}(t,r) &=& \frac{1}{A^2}\int g_{rr}Y^*_{\ell m}d\Omega
\\
G^{(\ell m)}(t,r) &=& \frac{1}{R^2 \ell(\ell+1)(\ell-1)(\ell+2)} \int
        \left\{\left(g_{\theta\theta}-\frac{g_{\phi\phi}}{\sin^2\theta}\right)
        \left(Y^*_{\ell m,\theta\theta}-\cot\theta Y^*_{\ell
        m,\theta}-\frac{1}{\sin^2\theta} Y^*_{\ell
        m,\phi\phi}\right)\right.
\\
        &&\left.  +\frac{4}{\sin^2\theta}g_{\theta\phi}(Y^*_{\ell
   m,\theta\phi}-\cot\theta Y^*_{\ell m,\phi}) \right\}d\Omega
\\
K^{(\ell m)}(t,r) &=& \frac{\ell (\ell +1)}{2}G^{(\ell m)}(t,r)
        +\frac{1}{2R^2}\int\left(g_{\theta\theta}+\frac{1}{\sin^2\theta}g_{\phi\phi}
        \right) Y^*_{\ell m} d\Omega
\end{eqnarray}

\section{Linear Solution for Conformal Factor}
\label{app:linsol}

In this appendix we derive an exact linear solution
for the perturbed black hole initial data sets used in this paper.
Expanding the conformal factor $\tilde{\psi}$  in terms of
the Brill wave amplitude $a$, we write
\begin{equation}
\tilde{\psi} = \tilde{\psi}^{(0)}+a \tilde{\psi}^{(1)} + {\cal O}(a^2)
\end{equation}
The zeroth order Hamiltonian constraint is then
\begin{equation}
\partial_\eta^2\tilde{\psi}^{(0)}+\partial^2_\theta\tilde{\psi}^{(0)}
+\cot\theta\partial_\theta \tilde{\psi}^{(0)}+\csc^2\theta
\partial^2_\phi\tilde{\psi}^{(0)}=0
\end{equation}
with the spherical Schwarzschild solution
\begin{equation}
\tilde{\psi}^{(0)} = \sqrt{2M} \cosh \left( \frac{\eta}{2}\right)
\end{equation}
Keeping terms linear in $a$, the first order Hamiltonian 
constraint is
\begin{equation}
4\partial_\eta^2\tilde{\psi}^{(1)}+4\partial_\theta^2\tilde{\psi}^{(1)}
+4\cot\theta\partial_\theta \tilde{\psi}^{(1)}+4\csc^2\theta\partial^2_\phi
\tilde{\psi}^{(1)}-\tilde{\psi}^{(1)}=
-\frac{1}{a}\tilde{\psi}^{(0)}(\partial^2_\eta q+\partial^2_\theta q
+ 2 \csc^2 \theta \partial^2_\phi q)
\label{1st_order_constraint}
\end{equation}
When $\tilde{\psi}^{(1)}$ is expanded in spherical harmonics,
\begin{equation}
\tilde{\psi}^{(1)} = \sum^\infty_{\ell=0} \sum^\ell_{m=-\ell} 
\tilde{\psi}^{(1)}_{\ell m} Y_{\ell m}
\label{1st_order_sum}
\end{equation}
Equation~\ref{1st_order_constraint} reduces to a series of ODEs, 
which can be solved for a particular case of the Brill wave function,
$q$. The solution for two of the Brill wave functions considered in this
paper are,

\begin{enumerate}

\item $q=2 a \sin^2\theta \exp(-\eta^2)$.

The only nonzero coefficients in the expansion~\ref{1st_order_sum} are 
$\tilde{\psi}^{(1)}_{00}$ 
and $\tilde{\psi}^{(1)}_{20}$ which have the values
\begin{eqnarray}
\tilde{\psi}^{(1)}_{00} &=& 
  -\frac{2\sqrt{2M\pi}}{3}e^{-\eta^2}\cosh(\eta/2)
  +\frac{2\pi}{3}[\cosh(\eta/2)+\mbox{erf}(\eta)\sinh(\eta/2)]
\\
\tilde{\psi}^{(1)}_{20} &=& 
\frac{2\sqrt{10M\pi}}{15} e^{-\eta^2} -\frac{2\sqrt{5}\pi}{15}e^{9/4}
\cosh(5\eta/2)+\sqrt{5\pi}{15}e^{9/4}\left[\mbox{erf}\left(-\eta+\frac{3}{2}
\right)e^{-5\eta/2}+\mbox{erf}\left(\eta+\frac{3}{2}\right)e^{5\eta/2}
\right]
\end{eqnarray}

\item $q=2 a \sin^4\theta e^{-\eta^2}(1+c \cos^2\phi)$

For this case, which includes a non-axisymmetric contribution through
the parameter $c$, there are several more modes in the expansion. 
The non-zero coefficients in~\ref{1st_order_sum} are now
$\tilde{\psi}^{(1)}_{00}$, $\tilde{\psi}^{(1)}_{20}$,  
$\tilde{\psi}^{(1)}_{22}\equiv \tilde{\psi}^{(1)}_{2,-2}$, 
$\tilde{\psi}^{(1)}_{40}$, 
$\tilde{\psi}^{(1)}_{42}\equiv\tilde{\psi}^{(1)}_{4,-2}$.

\begin{eqnarray}
\tilde{\psi}^{(1)}_{00} &=& \frac{2}{15} \sqrt{2\pi M} \left( 2+c \right)
\left[-\cosh\left(\frac{\eta}{2}\right) \left( 2 e^{-\eta^2} + \sqrt{\pi}
\right)+\sqrt{\pi} \sinh\left(\frac{\eta}{2}\right) \erf\left(\eta\right)
\right]
\\
\tilde{\psi}^{(1)}_{20} &=& \frac{1}{105} \sqrt{\frac{2 \pi M}{5}} \left(2+c
\right) \left[ 40 e^{-\eta^2} \cosh\left(\frac{\eta}{2}\right) + 6\sqrt{\pi} e
\cosh\left(\frac{5\eta}{2}\right) - 14 \sqrt{\pi} e^{9/4} \cosh\left(
\frac{5\eta}{2}\right) \right. \nonumber \\
&&\left. + 3 \sqrt{\pi} e^{1-5\eta/2} \erf\left(\eta-1\right) -
3\sqrt{\pi} e^{1+5\eta/2} \erf\left(\eta+1\right) - 7 \sqrt{\pi} e^{9/4-
5\eta/2} \erf\left(\eta-\frac{3}{2}\right) + 7 \sqrt{\pi} e^{9/4+5\eta/2}
\erf\left(\eta+\frac{3}{2}\right) \right]
\\
\tilde{\psi}^{(1)}_{22} &=& \frac{1}{70} \sqrt{\frac{\pi M}{15}} c \left[ -60
e^{-\eta^2} \cosh\left(\frac{\eta}{2}\right) - 44 \sqrt{\pi} e \cosh\left(
\frac{5\eta}{2}\right) - 14 \sqrt{\pi} e^{9/4} \cosh\left(\frac{5\eta}{2}
\right) \right. \nonumber \\
&&\left. - 22\sqrt{\pi} e^{1-5\eta/2} \erf\left(\eta-1\right) + 22\sqrt{\pi}
e^{1+5\eta/2} \erf\left(\eta+1\right) - 7\sqrt{\pi} e^{9/4-5\eta/2}
\erf\left(\eta-\frac{3}{2}\right) + 7\sqrt{\pi} e^{9/4+5\eta/2}
\erf\left(\eta+\frac{3}{2}\right) \right]
\\
\tilde{\psi}^{(1)}_{40} &=& -\frac{1}{105} \sqrt{2\pi M} \left(2+c\right) \left[
4 e^{-\eta^2} \cosh\left(\frac{\eta}{2}\right) - 2\sqrt{\pi} e^{25/4}
\cosh\left(\frac{9\eta}{2}\right) - \sqrt{\pi} e^{25/4-9\eta/2} \erf\left(
\eta-\frac{5}{2}\right) \right.  \nonumber \\
&& \left. + \sqrt{\pi} e^{25/4+9\eta/2} \erf\left(\eta+
\frac{5}{2}\right) \right]
\\
\tilde{\psi}^{(1)}_{42} &=& \frac{1}{42} \sqrt{\frac{\pi M}{5}} c \left[
4 e^{-\eta^2} \cosh\left(\frac{\eta}{2}\right) - 2\sqrt{\pi} e^{25/4}
\cosh\left(\frac{9\eta}{2}\right) - \sqrt{\pi} e^{25/4-9\eta/2} \erf\left(
\eta-\frac{5}{2}\right) \right. \nonumber \\
&& \left. + \sqrt{\pi} e^{25/4+9\eta/2} \erf\left(\eta+
\frac{5}{2}\right) \right]
\end{eqnarray}
If the parameter $c=0$, the 
coefficients with $m\neq 0$ vanish.
\end{enumerate}

Note that if the physical metric is required to linear order,
(as used in the examples in this paper), then the 
components will be given by, for example
\begin{eqnarray}
g_{\eta\eta} &=& \tilde{\psi}^4 e^{2q} 
\nonumber \\
&=& (1+2q)(\tilde{\psi}^{(0)})^4 + 4a(\tilde{\psi}^{(1)})^3+{\cal O}(a^2)
\end{eqnarray}
\section{Numerical Methods}

The numerical implementation of the 2D and 3D nonlinear codes
for the generation and evolution of initial data is described
fully in~\cite{Anninos94c,Bernstein93b,Anninos93c,Anninos94d}.

The procedure for extracting waveforms in both the nonlinear initial data, 
and in the evolved nonlinear data is the same.
The implementation consists of three main steps:

\begin{description}
  
\item{(i)} Find a 2-sphere on which the approximate spherical symmetry
  is manifest. For the examples used in this paper, appropriate
  spheres are known from the construction of the initial data, and the
  2-spheres are simply spheres of constant isotropic coordinate radius
  $\eta$ from the center of the octant symmetry.

\item{(ii)} Construct the spatial metric components $g_{\eta\eta}$,
  $g_{\eta\theta}$, $g_{\eta\phi}$, $g_{\theta\theta}$,
  $g_{\theta\phi}$, $g_{\phi\phi}$ on the given 2-sphere.  In
  general this procedure will involve interpolating metric components
  from the numerical grid used to create the initial data or 
  for the evolutions to a 2D numerical grid on the given 2-sphere, and
  then transforming the metric to the polar coordinate system. 
  However for the data sets used here, the
  initial data was created and evolved on the same 2-spheres used for the
  extraction. 

\item{(iii)} Integration over the 2-sphere to calculate the
  Regge-Wheeler functions, and directly from these the radiative
  variables $Q^+_{\ell m}$ and $Q^\times_{\ell m}$. The integrations 
  were performed numerically using Simpson's
  rule, which calculates the integral with an error of ${\cal
    O}(\Delta \theta^2 , \Delta \phi^2)$. To avoid complications, the
  polar grid points on the sphere are arranged so that the polar axis
  is straddled.  The number of polar grid points needed for an
  accurate estimate of the integral will depend in general on the
  angular structure of the metric components $g_{\alpha\beta}$ and 
  the $\ell$, $m$ mode under consideration. 

\end{description}


\begin{thebibliography}{10}

\bibitem{Anninos96c}
P. Anninos, J. Mass\'o, E. Seidel, and W.-M. Suen, Physics World {\bf 9},  43
  (1996).

\bibitem{Seidel98b}
E. Seidel,  in {\em Proceedings of the Fifteenth Meeting of the International
  Society of General Relativity and Gravitation}, edited by N. Dadich and J.
  Narlikar (IUCAA, Pune, India, 1998), in press.

\bibitem{Flanagan97b}
\'{E}. \'{E}.~Flanagan and S.~A. Hughes, Phys. Rev. D {\bf 57},  4566
  (1998).

\bibitem{Flanagan97a}
\'{E}. \'{E}.~Flanagan and S.~A. Hughes, Phys. Rev. D {\bf 57},  4535
  (1998).

\bibitem{Anninos94b}
P. Anninos, D. Hobill, E. Seidel, L. Smarr, and W.-M. Suen, Phys. Rev. D {\bf
  52},  2044  (1995).

\bibitem{Anninos94c}
P. Anninos, K. Camarda, J. Mass\'o, E. Seidel, W.-M. Suen, and J. Towns, Phys.
  Rev. D {\bf 52},  2059  (1995).

\bibitem{Camarda97b}
K. Camarda and E. Seidel, Phys. Rev. D {\bf 57},  R3204  (1998).

\bibitem{Gomez98a}
R. Gomez {\it et~al.}, Phys. Rev. Lett. {\bf 80},  3915  (1998).

\bibitem{Brandt94a}
S. Brandt and E. Seidel, Phys. Rev. D {\bf 54},  1403  (1996).

\bibitem{Brandt94b}
S. Brandt and E. Seidel, Phys. Rev. D {\bf 52},  856  (1995).

\bibitem{Brandt94c}
S. Brandt and E. Seidel, Phys. Rev. D {\bf 52},  870  (1995).

\bibitem{Abrahams92a}
A. Abrahams, D. Bernstein, D. Hobill, E. Seidel, and L. Smarr, Phys. Rev. D
  {\bf 45},  3544  (1992).

\bibitem{Bernstein93b}
D. Bernstein, D. Hobill, E. Seidel, L. Smarr, and J. Towns, Phys. Rev. D {\bf
  50},  5000  (1994).

\bibitem{Anninos93b}
P. Anninos, D. Hobill, E. Seidel, L. Smarr, and W.-M. Suen, Phys. Rev. Lett.
  {\bf 71},  2851  (1993).

\bibitem{Baker96a}
J. Baker, A. Abrahams, P. Anninos, S. Brandt, R. Price, J. Pullin, and E.
  Seidel, Phys. Rev. D {\bf 55},  829  (1997).

\bibitem{Price94b}
P. Anninos, R.~H. Price, J. Pullin, E. Seidel, and W.-M. Suen, Phys. Rev. D
  {\bf 52},  4462  (1995).

\bibitem{Pullin98a}
J. Pullin,  in {\em Proceedings of the Fifteenth Meeting of the International
  Society on General Relativity and Gravitation}, edited by N. Dadich and
  J. Narlikar (IUCAA, Pune, India, 1998), in press.

\bibitem{Masso98a}
C. Bona, J. Mass\'o, E. Seidel, and P. Walker, gr-qc/9804065, submitted to
  Phys. Rev. D.

\bibitem{Seidel92a}
E. Seidel and W.-M. Suen, Phys. Rev. Lett. {\bf 69},  1845  (1992).

\bibitem{Cook97a}
G.~B. Cook {\it et~al.}, Phys. Rev. Lett {\bf 80},  2512  (1998).

\bibitem{Gomez97a}
R. Gomez, L. Lehner, R. Marsa, and J. Winicour, Phys. Rev. D {\bf 57},  4778
  (1998).

\bibitem{Gomez97b}
R. Gomez, R. Marsa, and J. Winicour, Phys. Rev. D {\bf 56},  6310  (1997).

\bibitem{Price94a}
R.~H. Price and J. Pullin, Phys. Rev. Lett. {\bf 72},  3297  (1994).

\bibitem{Regge57}
T. Regge and J. Wheeler, Phys. Rev. {\bf 108},  1063  (1957).

\bibitem{Moncrief74}
V. Moncrief, Annals of Physics {\bf 88},  323  (1974).

\bibitem{Gerlach79}
U. Gerlach and U. Sengupta, Phys. Rev. D. {\bf 19},  2268  (1979).

\bibitem{Seidel90c}
E. Seidel, Phys. Rev. D {\bf 42},  1884  (1990).

\bibitem{Abrahams88}
A. Abrahams, Ph.D. thesis, University of Illinois, Urbana, Illinois, 1988.

\bibitem{Abrahams89}
A. Abrahams,  in {\em Frontiers in Numerical Relativity}, edited by C. Evans,
  L. Finn, and D. Hobill (Cambridge University Press, Cambridge, England,
  1989).

\bibitem{Abrahams90}
A. Abrahams and C. Evans, Phys. Rev. D {\bf 42},  2585  (1990).

\bibitem{Bernstein94a}
D. Bernstein, D. Hobill, E. Seidel, and L. Smarr, Phys. Rev. D {\bf 50},  3760
  (1994).

\bibitem{Brandt97a}
S. Brandt, K. Camarda, and E. Seidel, in preparation.

\bibitem{Camarda97a}
K. Camarda, Ph.D. thesis, University of Illinois at Urbana-Champaign, Urbana,
  Illinois, 1998.

\bibitem{Brill59}
D.~S. Brill, Ann. Phys. {\bf 7},  466  (1959).

\bibitem{Camarda97c}
K. Camarda and E. Seidel, gr-qc/9805099.

\bibitem{Abrahams95b}
A. Abrahams and R. Price, Phys. Rev. D {\bf 53},  1963  (1996).

\bibitem{gleiser96a}
R.~J. Gleiser, C.~O. Nicasio, R.~H. Price, and J. Pullin, Class. Quant. Grav.
  {\bf 13},  L117  (1996).

\bibitem{Gleiser96b}
R.~J. Gleiser, C.~O. Nicasio, R.~H. Price, and J. Pullin, Physical Review
  Letters {\bf 77},  4483  (1996).

\bibitem{Allen98a}
G. Allen, K. Camarda, and E. Seidel,  in preparation.

\bibitem{Anninos93c}
P. Anninos, D. Bernstein, D. Hobill, E. Seidel, L. Smarr, and J. Towns,  in
  {\em Computational Astrophysics: Gas Dynamics and Particle Methods}, edited
  by W. Benz, J. Barnes, E. Muller, and M. Norman (Springer-Verlag, New York,
  1997), in press.

\bibitem{Anninos94d}
P. Anninos, J. Mass\'o, E. Seidel, W.-M. Suen, and M. Tobias, Phys. Rev. D {\bf
  56},  842  (1997).

\end{thebibliography}

\end{document}